Title:

# deepPERFECT: Novel Deep Learning CT Synthesis Method for Expeditious Pancreatic Cancer Radiotherapy

Running title:

Expeditious Pancreatic Cancer Radiotherapy

Authors and Affiliations:

Hamed Hooshangnejad[1,2,3], Quan Chen[4], Xue Feng[5], Rui Zhang[6], Kai Ding[2,3]

[1]Department of Biomedical Engineering, Johns Hopkins School of Medicine, Baltimore, MD, USA,

[2]Department of Radiation Oncology and Molecular Radiation Sciences, Johns Hopkins School of Medicine, Baltimore, MD, USA

[3]Carnegie Center of Surgical Innovation, Johns Hopkins School of Medicine, Baltimore, MD, USA

[4]City of Hope Comprehensive Cancer Center, Duarte, CA, USA

[5]Carina Medical LLC, Lexington, KY, USA

[6]Department of Surgery, University of Minnesota, Minneapolis, MN, USA

Corresponding Author:

Kai Ding

Department of Radiation Oncology and Molecular Radiation Sciences, Johns Hopkins School of Medicine, 401 N Broadway, Suite 1440, Baltimore, MD 21231-2410, USA.

Tel: 410-955-7391;

Fax: 410-502-1419;

Email: kding1@jhmi.edu

# Abstract

Background:

Pancreatic cancer is a devastating disease with more than 60,000 new cases each year and less than 10 percent 5-year overall survival. Radiation therapy (RT) is an effective treatment for Locally advanced pancreatic cancer (LAPC). The current clinical RT workflow, however, is lengthy and involves separate image acquisition for diagnostic CT (dCT) and planning CT (pCT) which imposes a huge burden on patients and their caretakers. Moreover, studies have shown a reduction in mortality rate from expeditious radiotherapy treatment course. Although, in theory, dCT can be used for RT planning, the differences in the image acquisition setup and patient's body demand a new scan to be acquired.

Purpose:

To address this issue, we are presenting deepPERFECT: deep learning-based Planning External-beam Radiotherapy Free from Explicit simCT, that adapts the shape of the patient body on dCT to the treatment delivery setup. Our method expedites the treatment course by allowing the design of the initial RT planning before the pCT acquisition. Thus, the physicians can evaluate the potential RT prognosis ahead of time, verify the plan on the treatment day-one CT and apply any online adaptation if needed.

Methods:

We used the data from 25 pancreatic cancer patients undergone stereotactic body radiation therapy. Each patient had a pair of dCT and pCT as part of their treatment course. The model was trained on 15 cases and tested on the remaining ten cases. We evaluated the performance of four different deep-learning architectures for this task. The synthesized CT (sCT) and regions of interest (ROIs) were compared with ground truth (pCT) using Dice similarity coefficient (DSC) and Hausdorff distance (HD). Finally, we evaluated the RT plan dose distribution for four scenarios.

Results:

We found that the three-dimensional Generative Adversarial Network (GAN) model trained on large patches has the best performance. Using the in-place deformed image identity loss enhanced the performance of the deepPERFECT in predicting the body shape. The synthesized deformation fields and CT scans were evaluated using multiple figures of merit. The average

DSC and HD for body contours were 0.93, and 4.6 mm. Additionally, we evaluated the quality of clinical-grade radiotherapy plans designed using the synthesized CT, by comparing the dosimetric indices measured on synthesized CT and ground truth. We found no statistically significant difference between the synthesized CT plans and the ground truth.

Conclusions:

We showed that deepPERFECT predicts the shape of the patient body on pCT using the dCT scan with good performance. We believe employing deepPERFECT shortens the current lengthy clinical workflow by at least one week and improves the effectiveness of treatment and the quality of life of pancreatic cancer patients.

## Introduction

Pancreatic cancer is a devastating disease with more than 60,000 new cases each year and less than 10% 5-year overall survival rate[1]. Pancreatic cancer patients are at great risk of distant progression, so achieving local control (LC) is critical for these patients[1–6]. Radiation therapy (RT) is an effective treatment for achieving LC[7–10]. Rapid treatment planning and delivery are critical for aggressive pancreatic cancer and a week delay increases the possibility of spreading or recurring[11,12], yet the current RT workflow is considerably time-consuming.

Figure 1 shows the current RT workflow which consists of numerous steps. Multiple appointments and image acquisition result in significant wait time, and a huge burden on patients and caretakers[13,14]. One major source of delay in the workflow is due to the several appointments and separate image acquisitions, namely acquiring a diagnostic computed tomography (dCT) scan and the planning CT (pCT) scan. This results in a median of 15 days delay for patient diagnosis and 15 more from diagnosis to treatment initiation[15].

Reducing the delays to treatment initiation has a significant clinical impact. It has been shown that shorter therapy initiation is associated with improved survival[16]. Amongst more than 29,000 patients, starting any treatment within 6 weeks improved median OS[16]. Moreover, the data from more than 70,000 pancreatic cancer patients shows the most substantial associations with worsened mortality were seen among other cancer types with a 3.2% increase in mortality per week of delay[17]. Moreover, reducing the time before surgery to 32 days reduces the risk of the progression of the tumor to the unresectable stage by half compared with a longer waiting time.

Because of the dCT's superior resolution, it is used for delineating the tumor. It is shown that the dCT can be used for RT planning[13,14,18,19]. However, in practice, because of the differences in image acquisition settings, dCT is not used for RT planning and a separate planning CT (pCT) is acquired, which results in a considerable delay between diagnosis and treatment delivery. In the case of the patients included in this study, there was a median of 11 days, ranging from 2 to 30 days delay between diagnostic scan and simulation scan. The previous attempts such as STAT RT[20] for using the day 1 onboard imaging for treatment planning before subsequent treatment delivery for palliative cases is not suitable for pancreatic cancer SBRT. The physicians need time on deciding the trade-off between target coverage and OAR sparing, and then the dose prescription.

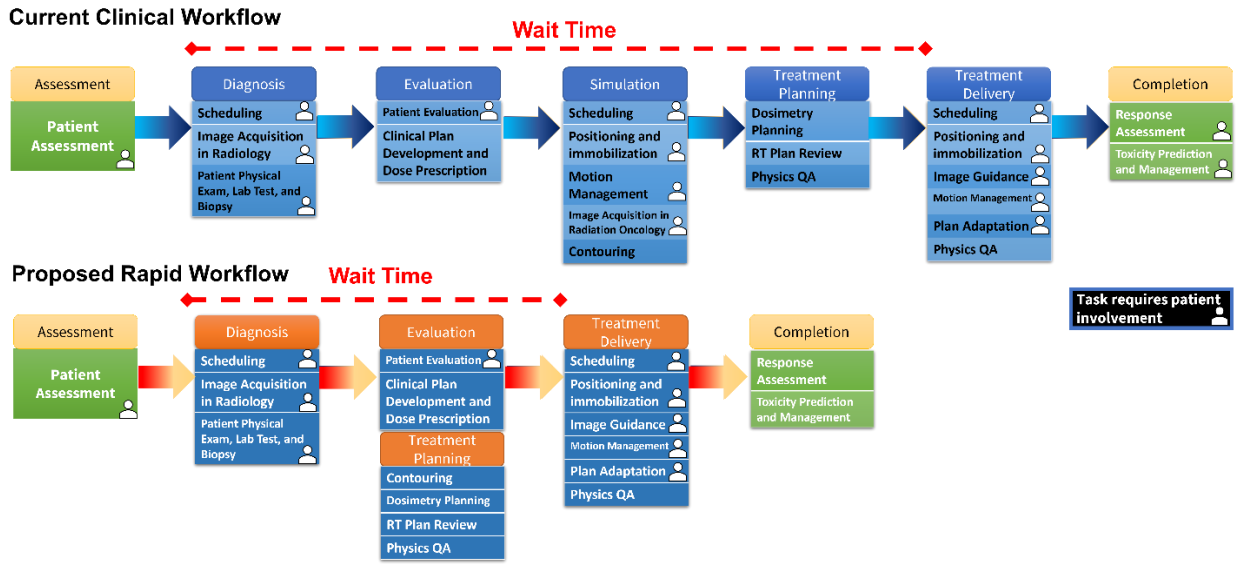

*Figure 1. the illustration of the current clinical workflow that due to the many steps before treatment results in considerable treatment delay. On the contrary, by using the synthesized planning CT, we proposed a rapid workflow that significantly reduces the treatment delay.*

Figure 2 shows a typical pair of dCT and pCT acquired in our institution. The use of a different couch in the radiology department and radiation oncology department results in a clear difference in the shape of the patient's back. In radiology CT scans, the curved couch top is used for patient comfort. By contrast, radiation oncology flat couch top focuses on the daily reproducibility of patient position. Also, to reduce patient movement, the active breath-hold technique is used as the standard acquisition procedure for pCT and treatment delivery in our institution, which causes a difference in patient anatomy on the two scans[21,22]. Thus, the design of the initial RT plan is only possible until after the acquisition of the pCT scan which results in a considerable stall in the treatment workflow.

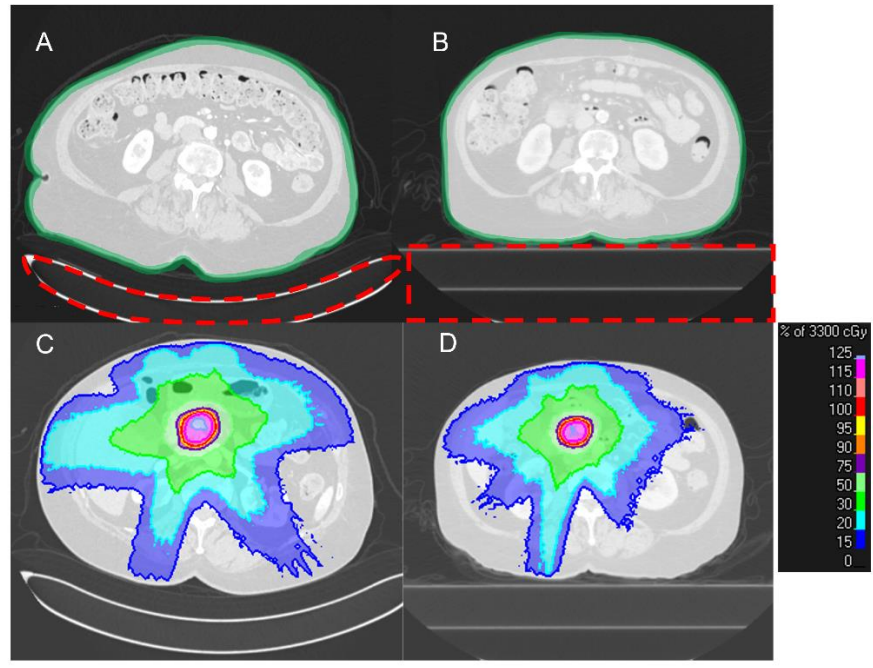

*Figure 2. A typical pair of dCT (A) and pCT (B) of a patient. (A) the difference in the shape of the couch results in a clear change in the patient's back curvature. Also, the overall shape of the body is different due to the active breath-hold motion management procedure that is used for pCT acquisition and treatment delivery. (C, D) the differences in patient body shape results in a difference in dose distribution (the couches are already removed for planning).*

To address the delay between diagnosis and treatment delivery, we developed deepPERFECT; a generative adversarial network (GAN) deep learning (DL) model. The adversarial loss that encourages the generation of data indistinguishable from real data, resulted in the huge success of GAN models in image synthesis applications[23], such as synthesizing CT from cone-beam CT [24,25], and magnetic resonance image[26]. deepPERFECT synthesizes the four-dimensional (4D) including 3 channels of 3D dimensional deformation fields (DF) that transform dCT to synthesized pCT (sCT) that can be used for initial verification of the RT treatment plan, thus, expediting the treatment course.

In our feasibility study, we showed that deepPERFECT can expedite the current clinical workflow and treatment course by removing the need for acquiring planning CT before designing the initial treatment RT plan. Given that there is a significant relationship between the increase in mortality and delay in treatment delivery[27], we believe deepPERFECT improves the quality of life for much-needed pancreatic cancer patients. More importantly, it allows the physicians to evaluate the potential RT prognosis ahead of time, and verify the plan on the treatment day-one CT. On Day 1 treatment CT, if necessary, the sCT plan will be adapted to the treatment delivery patient setup using on/off table imaging for same-day online ART such as on-table adaptive therapy methods like Ethos or same-day off-table online ART using Raystation adaptive treatment planning (RaySearch Laboratories, Sweden).

## Methods and Materials

### Data Preparation

We included the data from 25 pancreatic cancer patients, treated with stereotactic body radiation therapy (SBRT) in our institution under Internal Review Boards (IRBs) approval (15 cases for training and validation, and ten cases for testing and RT planning). The dCT and pCT scans were acquired with 120 KVp, 200 mA, and 50 cm field of view. dCT scans have slice thickness ranging from 0.5 to 1.25 mm, and pCT scans' slice thickness is 2 mm. To have a uniform 3D physical dimension, in practice, all scans can be resampled to 1 mm slice thickness, but due to GPU limitation, we resampled to 2.5 mm. The pCT was contoured by physicians as part of the standard of care in our institution, and the dCT and sCT contours were generated by applying the deformation vector fields to dCT contours, and then, were verified by a physician.

Using an in-house robust couch removal algorithm, we, first, removed the couch from the scans, only keeping the patient's body. Because the treatment couch is identical, it was later digitally added to the synthesized scans. pCT and dCT were, first, aligned by the spine. Because pCT is not yet acquired at the time of dCT acquisition, we chose the dCT as the physical origin for rigid registration and used the ROI-restricted rigid registration using an in-house automatic spine segmentation algorithm, to further align the spines.

Next, the dCT was deformably registered to pCT. To avoid the unreal deformation of the spine, such as elongation of vertebrae and disk, we used the spine mask to define a rigidity penalty term. Because the high contrast lung region dominated the result of the registration, we performed a two-step sequential registration, first, the whole body was registered, and automatically segmented lungs were excluded for further abdominal-focused registration. The abdominal mask was determined by using the lowest point on lung contours while keeping the spine rigid penalty to avoid unreal deformation. All registrations were done by the state-of-the-art Elastix image registration algorithm[28,29].

As 15 cases were too few for successful training of a DL model, we augmented the data by applying random shifts of between -20 mm to 20 mm and random rotations of -10 to 10 degrees along the z-axis (depth) and lastly, we used randomly generated patches of the CT images as the input to the model. Overall, in each epoch, the DL model was trained on more than 6000 augmented data.

## Deep Learning Model

We used four network configurations: (1) 3D U-Net Convolutional network with a patch size of(128*128*128), (2) 2.5D Pix2Pix generative adversarial network (GAN) with three adjacent slices of (128*128), (3) 3D Pix2Pix GAN with small patch size (32*32*32), and (4) 3D Pix2Pix GAN with large patch size (128*128*128). The output of the DL model is the DFs. We chose to generate DFs rather than CT scans to keep the CT intensity calibration intact. The intensity of dCT and pCT is based on careful calibration and quality assurance of the CT scanner and is crucial for accurate dose calculation. Because DFs are real numbers with positive and negative values, we used leaky rectified linear unit activation function $f(x) = \max(0.01x, x)$. Batch normalization was applied to the output of convolutional layers. deepPERFECT was trained with three loss functions: (1) DF identity loss, the L1-norm of the difference between the true and generated DFs, (2) abdominal identity loss, the L1-norm of difference in intensity of the abdominal portion of pCT and intermediate sCT scan, created using the generated DFs applied to dCT in-place, (3) the adversarial loss for GAN architecture. A weighted Adam optimizer was used to optimize the loss functions. Figure 3 shows an overview of the DL model and loss used for training.

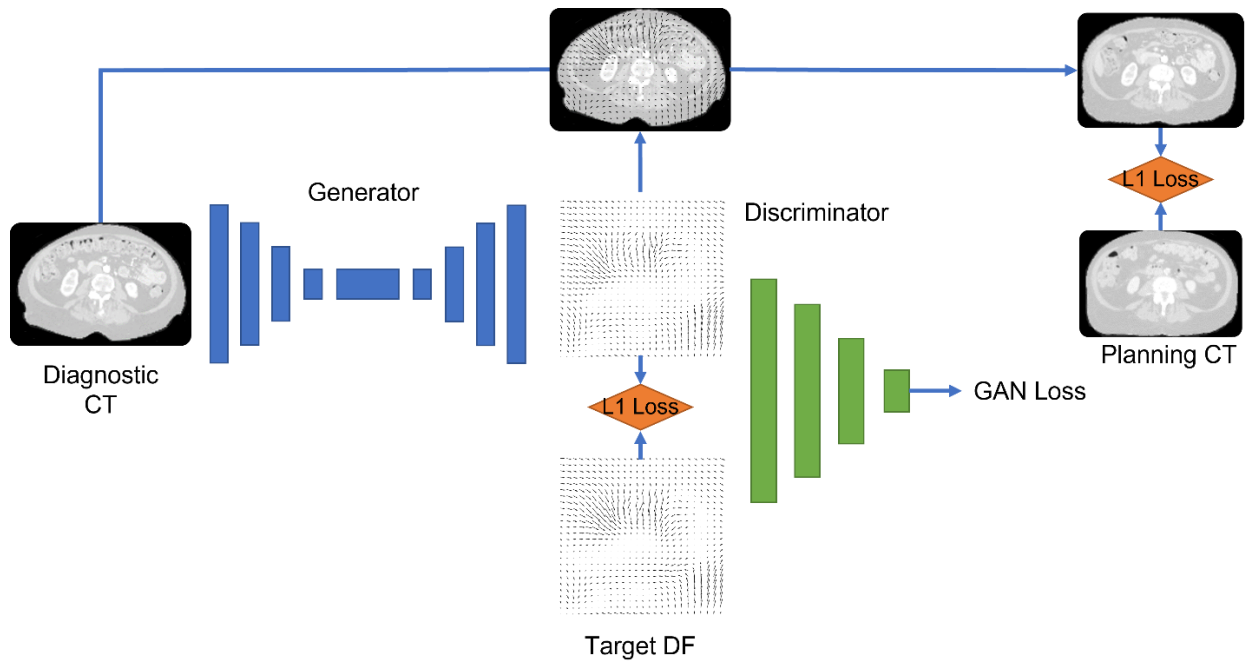

*Figure 3. An overview of the 3D GAN model and losses used to train the model.*

## Training and Testing of the Model

Out of 15 cases for training, we used data from 13 patients for training and 2 for validation. After the hyperparameters are set, we trained the model on all 15 cases and tested it on the 10 leave-out cases. We used the following cost function for deepPERFECT training.

$$loss = \lambda_1 L_{GAN}(G,D) + \lambda_2\, L_{L1}(G) + \lambda_3\, L_{L1}(I) + \lambda_4\, R_{smooth}(G)$$

$\lambda$ is the hyparameter that indicates the effect of each loss on the final loss function value. In this equation, G denotes the generator output which is deformation vector fields and D denotes the discriminator. $I$ denotes the final deformed image and R represents the regulatory term. $L_{GAN}(G,D)$ is the adversarial loss defined as:

$$L_{GAN}(G,D) = E_{x,y}\,(\log D(x,y)) + E_x(\log\left(1 - D(x,G(x))\right))$$

In which x represents the input and y is the target deformation vector field.

$L_{L1}(G)$ is the L1 norm of the differences between target DVFs and network generator DVFs defined as:

$$L_{L1}(G) = E_{x,y}(|y - G(x)|)$$

$L_{L1}(I)$ is the L1 norm between target image $I$ and deformed image, created by applying G(x) or $G_x$ to the input image, defined as:

$$L_{L1}(I) = E_{x,y}(|I - G_x(x)|)$$

Finally, to enforce smooth deformation fields, we used the second-order curvature regulatory term, widely used in registration literature, given by

$$R_{smooth}(G) = \int \sum_{j=1}^{3} \|\Delta G_i(x)\|^2$$

## Radiation Therapy Planning

We planned the ten test cases with volumetric modulated arc therapy (VMAT) SBRT (33Gy in 5 fx), according to the pancreatic SBRT planning protocol in our institution. The planning target volume (PTV) was created by a 2 mm expansion of mock multiple active breath-hold (GTV-multabh), which itself is a 3mm uniform expansion of GTV. For further details please refer to our previous studies[5,21,22]. The target objectives for RT planning were as follows: 100% of PTV receive 25 Gy, at least 95% of PTV volume receive 33 Gy, less 1cc volume of PTV receives more than 42.9 Gy, 100% of GTV receive 33 Gy, at least 95% of GTV-multabh receive 33 Gy. The organs at risk (OAR) constraints were as the following: less than 20 ccs of the bowel, duodenum, and stomach receive 20 Gy, less than 1cc of the bowel, duodenum, and stomach receive 33 Gy, and less than 25% of kidney receive 12 Gy, less than 50% of liver receive 12 Gy, and less than 1 cc of the spinal cord receive 8 Gy. We used Raystation (RaySearch Laboratories, Stockholm, Sweden) treatment planning system for plan optimization and dose distribution calculation.

As shown in Figure 4, for each patient, we designed and optimized two VMAT plans on pCT and sCT. The dose distribution was calculated for four scenarios. (1) the dose distribution of pCT plan on pCT scan (the ground truth dose distribution). (2) the dose distribution of sCT plan on sCT scan. (3) the "Plan Recalculation" scenario in which we recalculated the dose distribution of sCT plan on pCT scan using the pCT isocenter by shifting the iso-center of the sCT plan beams to pCT iso-center. Then, we evaluate the dosimetric indices using the pCT contours. (4) the "Synthesized ROIs on Planning CT" scenario in which we recalculated the dose distribution of sCT plan on pCT scan using the sCT isocenter (no shift in isocenter). Then, we evaluate the dosimetric indices using the sCT contours mapped to the pCT. Using this scenario, we assured that the observed difference between ground truth and the "Plan Recalculation" scenario is due to the unpredictable and patient specific abdominal anatomical variation and not the CT quality.

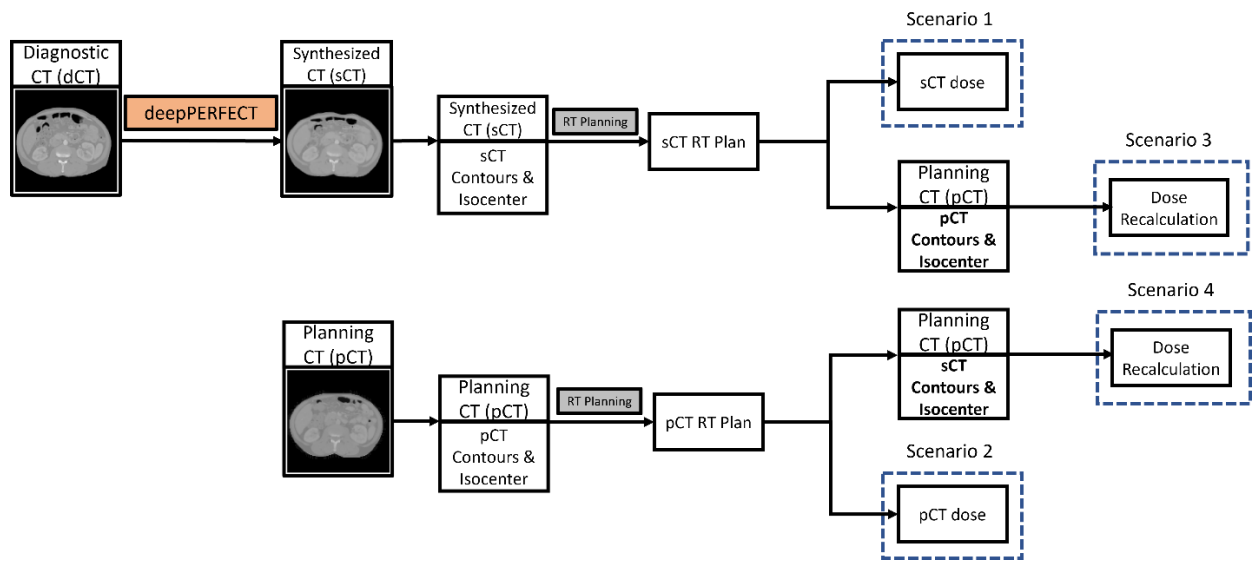

Figure 4. Illustration of the RT planning scenarios.

## Evaluation metrics

First, we evaluated deepPERFECT by comparing the intensity of pCT scans with sCT scans using the root averaged squared sum of differences (RASSD). Secondly, the body contours on the two scans were compared using the Dice similarity coefficient (DSC), and Hausdorff distance (HD). No center of mass alignment was done for DSC and HD calculation. Lastly, we reported dose volume histogram point measurements, the V22 Gy and V33 Gy defined as the volume of the ROI receiving 22 Gy and 33 Gy for OARs (Duodenum, stomach, and bowel), and the percentage of target coverage with the prescribed dose (V33 Gy) for target volumes (GTV and PTV) for the four dose distribution scenarios, explained in the previous section.

## Statistical Analysis

Using the pairwise permutation test (n=10,000), we tested the equivalency of dosimetric indices. The normality assumption was circumvented by using a non-parametric permutation test.

## Placement of the Virtual Couch

To synthesize the sCT, the couch is first removed from the dCT scans, but the treatment couch is required for accurate dose calculation, Thus, later in the process, we developed an algorithm that augments the sCT with the couch by placing a virtual couch with its surface tangent to the back of the patient.

# Results

Figure 5 shows the results for an example test case. Figure 5 (A) shows dCT (input to the model), (B) pCT (ground truth), and (C-F) the sCT for 3D Pix2Pix with large patches, U-Net, 2.5D Pix2Pix, and 3D Pix2Pix with small patches. As seen, the back of the patient and the overall shape of the abdominal area have the most similarity to sCT generated by 3D Pix2Pix with large patches (Figure 5(C)).

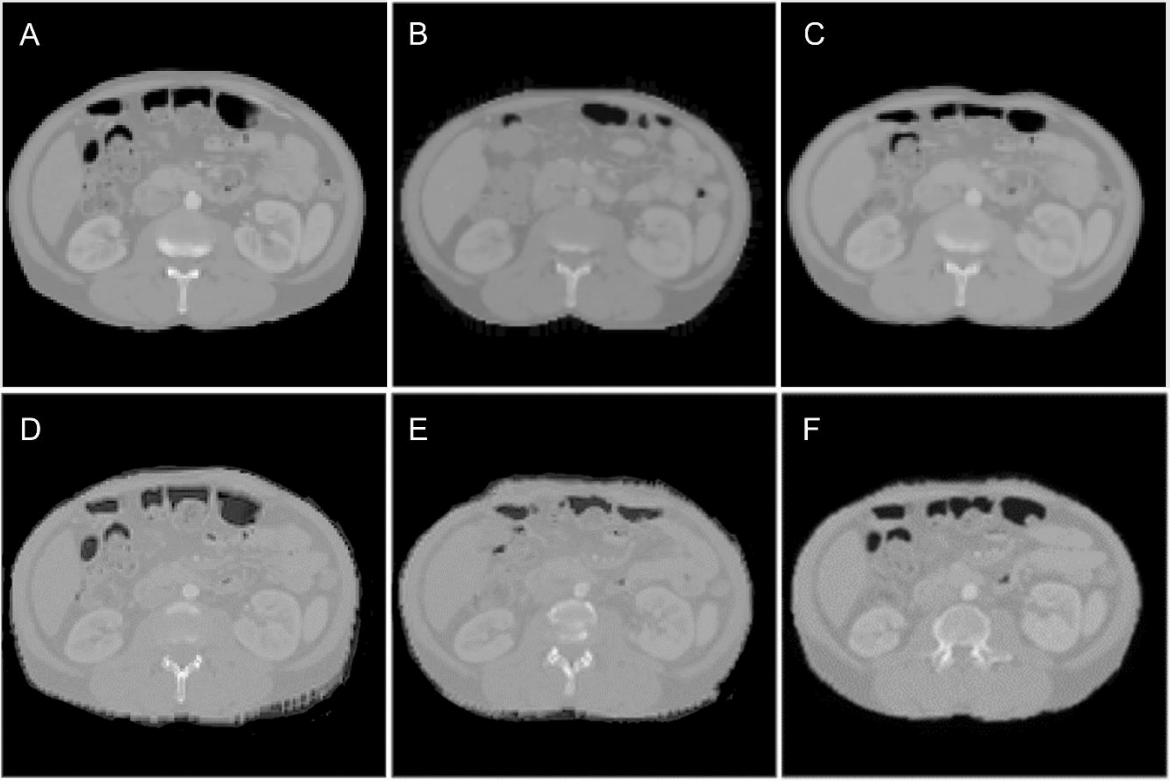

*Figure 5. An illustration of planning CT synthesis result: (A) the dCT, the input to the model, (B) the pCT, the ground truth, and (C-F) the sCT generated by Pix2Pix 3D large patches, U-Net, Pix2Pix 2.5. and Pix2Pix small patches, respectively.*

We evaluated the quality of the sCT scan using the ASSD, DSC, and HD for the entire body contour. The result is summarized in table 1. The 3D Pix2Pix trained on the large patches showed the best performance among the other configurations. We used this configuration to generate sCT for the RT planning part of the study.

*Table 1. The result of the image quality evaluation*

| Model Architecture | Pix2Pix 3D Large Patch | Pix2Pix 3D Small Patch | Pix2Pix 2.5D | U-Net |
|---|---|---|---|---|
| Metric | Average ± STD | | | |
| RASSD (HU) | **334 ± 65** | 541 ± 83 | 874 ± 156 | 1242 ± 132 |
| DSC body contour | **0.93 ± 0.04** | 0.81 ± 0.12 | 0.62 ± 0.09 | 0.57 ± 0.08 |
| HD body contour (mm) | **4.6 ± 2.1** | 15.2 ± 5.9 | 28.1 ± 6.2 | 35.7 ± 6.7 |
| DSC GTV | **0.82 ± 0.12** | 0.69 ± 0.13 | 0.64 ± 0.16 | 0.61 ± 0.13 |
| HD GTV (mm) | **7.12 ± 3.1** | 14.8 ± 8.9 | 18.3 ± 9.2 | 25.8 ± 8.6 |

For the best model, Pix2Pix 3D with large patches, the average and standard deviation difference in GTV volume on pCT and sCT was 1.1±1.8 cc. The GTV minimum distance to the main OARs namely duodenum, stomach, and bowel was also measured. The average and standard deviation of the difference in minimum distance between GTV-duodenum, GTV-stomach, and GTV-bowel were 0.37±1.1 mm, 0.52±1.4 mm, 0.61± 1.5 mm, respectively. There was a median of 11 days, ranging from 2 to 30 days delay between the diagnostic scan and simulation scan, and a median of 8 days to treatment initiation. The averaged DSC between dCT and pCT was 0.62 ± 0.1, which increased to 0.82 ± 0.12 for pCT and sCT.  Moreover, the DSC Figure 6 shows the diagnostic CT (A), planning CT (B), and synthesized CT (C) in the abdominal level/window (40/400) on the first row. The second row (D) is illustrating the HU intensity difference map between planning CT and diagnostic CT (pCT – dCT) and (E) between planning CT and synthesized CT (pCT – sCT). deepPERFECT predicted synthesized CT from diagnostic CT shows a high similarity to planning CT.

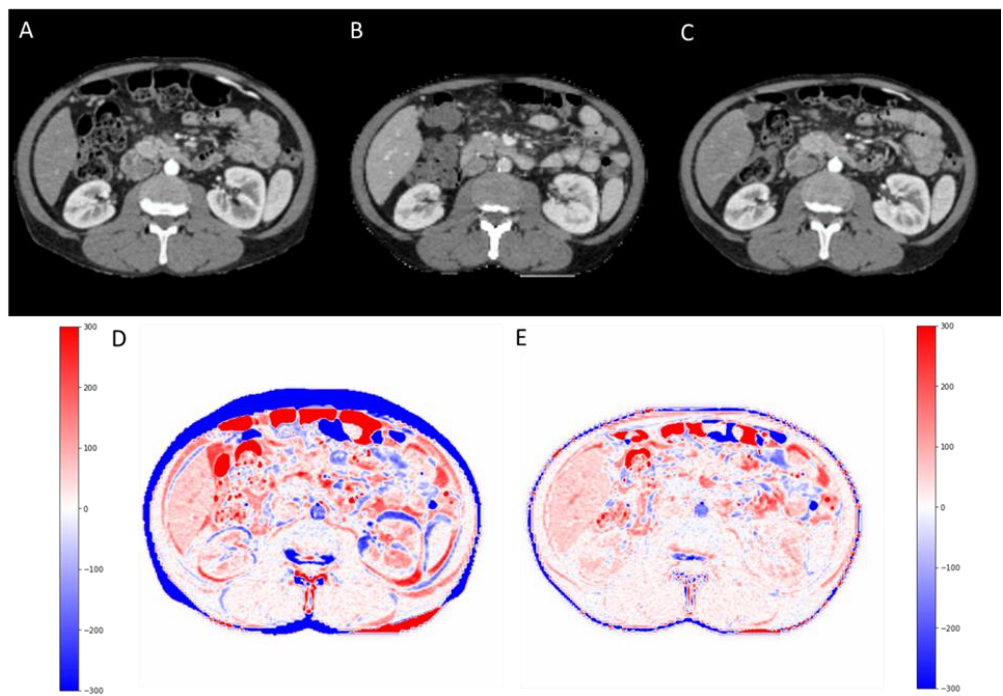

*Figure 6. First row: The diagnostic CT (A), planning CT (B), and synthesized CT (C) in abdominal level/window (40/400). Second row: (D) is illustrating the HU intensity difference map between planning CT and diagnostic CT (pCT – dCT) and (E) between planning CT and synthesized CT (pCT – sCT). deepPERFECT predicted synthesized CT from diagnostic CT shows a high similarity to planning CT.*

Next, we compared the SBRT plans using the OARs (duodenum, stomach, bowel) V20 Gy and V33 Gy, Dmax (max dose), and V100%, V95% for target volumes (GTV and PTV). The duodenal V33Gy had a marginally significant difference (p-value=0.049) between ground truth and plan recalculation. No other statistically significant differences were found in V20Gy and V33Gy. As expected, due to the unpredictable and patient-specific abdominal anatomical variations the "plan recalculation" scenario, in which the isocenter is only shifted, showed on average a 2% reduction in PTV coverage with the prescribed dose (33 Gy). PTV V95% and GTV V100%, and V95% were comparable. Figure S1 of supplementary material, shows the V95% coverage for GTV and PTV, as well as the Dmax for OARs.

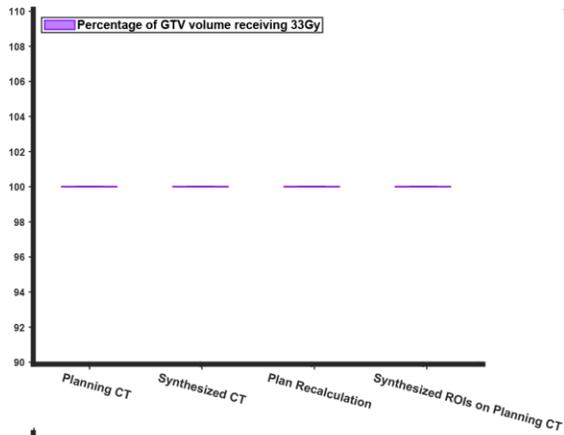
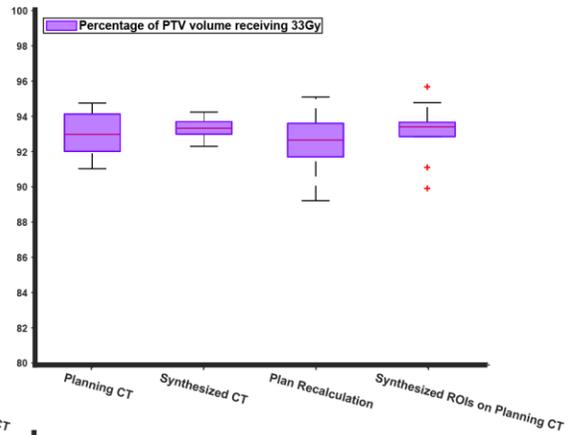
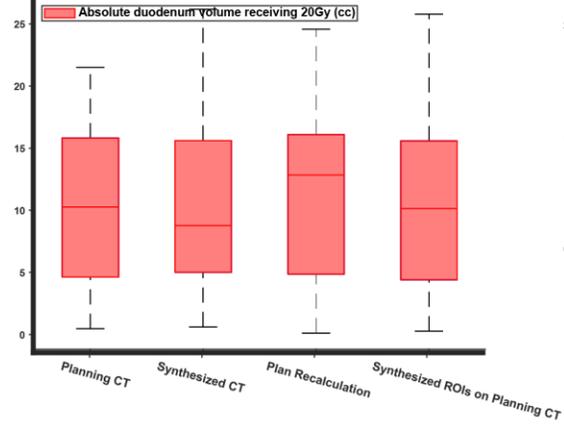
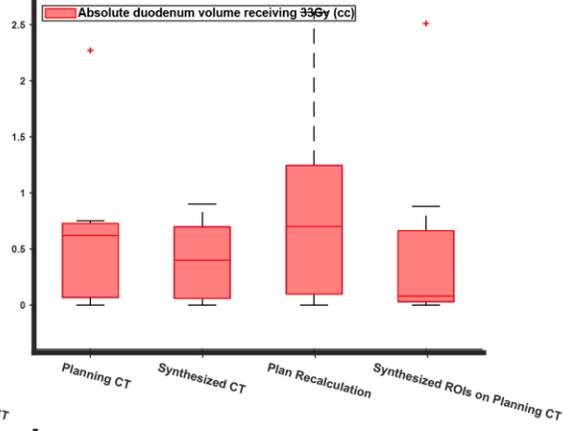
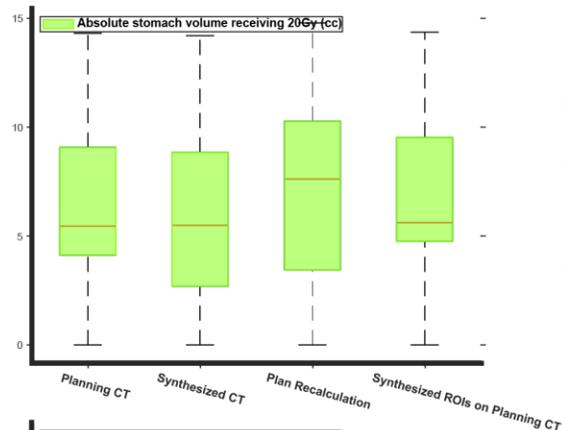
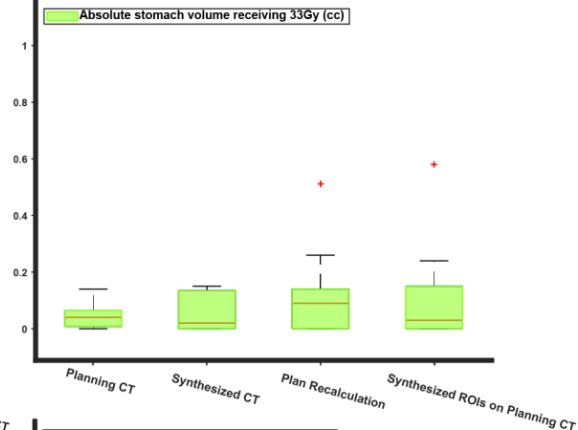
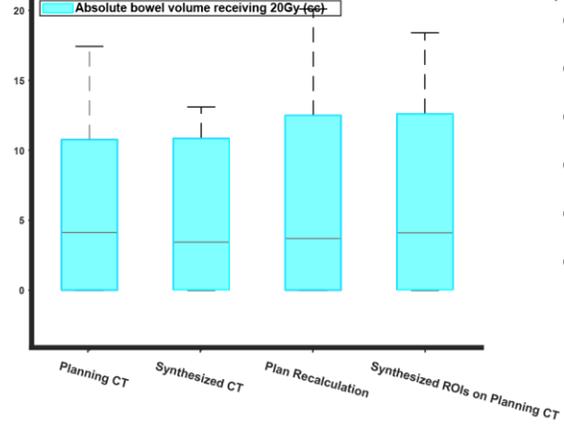
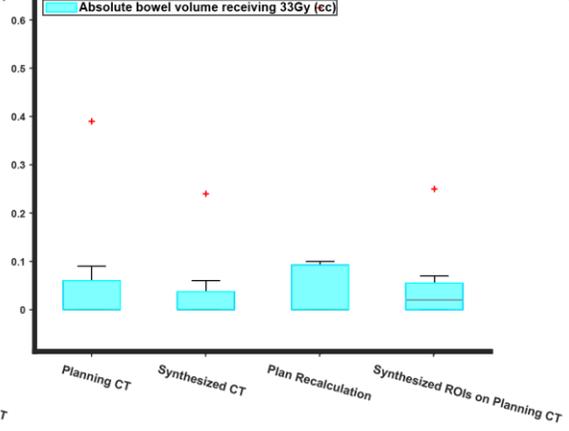

*Figure 6. Comparison of dosimetric indices for SBRT plans: the comparison was done for V33Gy for target volumes (GTV and PTV) and V33Gy and V20Gy for the proximal OARs (duodenum, stomach, and bowel). Each figure consists of four box plots for RT plans designed on pCT (ground truth), sCT, and plan recalculation (dose recalculation by shifting the isocenter of beams to pCT isocenter), Synthesized ROIs on planning CT (dose recalculation on pCT using sCT ROIs).*

## Discussion

We presented a novel DL system to reduce the extended delay before treatment delivery due to the separate scan acquisitions. Our method makes the dCT compatible with the treatment room setup, and thus, allows the initial RT plan to be designed. Thus, the physicians can evaluate the potential RT prognosis ahead of time, verify the plan on the treatment day-one CT and apply any online adaptation if needed. This reduces the wait time before the start of the treatment course. Our method can reduce the length of the treatment course by at least one week and given that the treatment delay increases mortality[27], we believe our method improves the patient's quality of life.

The source of delay from diagnosis to treatment delivery is not only the scheduling, wait time and acquision of planning CT, but also the treatment design, RT planning, verification. The physicians need time to make decision regarding the trade off between taget coverage and OAR sparing, and dose prescription. With treatment Day1, there is insufficient time to make a decision. For example, the patient may have previous RT treatment, or neoadjuvant or concurrent/adjuvant other cancer treatment therapies such as chemotherapy. All these are delayed until pCT acquisition, therefore syntheising the deepPERFECT generated sCT can provide enough time for physician to have well ahead of time assessment. With implementation of deepPERFECT, as soon as the diagnostic CT is available, the physicians can also evaluate the early prognosis of RT treatment, and decide on prescribtion, dose constraint, and inititatie the treatment.

In this study, we used the 3D Unet GAN architecture for developing the deepPERFECT framework. The 3D UNet GAN architecture has been long used for various medical image synthesis applications. It has been used for synthesizing PET scans from MRI[30,31], and synthesizing the CT scan from MRI[32] and frequently used in MRI reconstruction[33–37], low dose CT denoising[33,34,38,] optimization of the pre-trained network for sharpness detection and highlighting low contrast region in CT image[38], and many other applications. Due to 3D Unet GAN robust performance and versatility, we also based deepPERFECT design upon 3D Unet architecture. More advanced DL methods has also been used for similar applications that warrants future study to determine if they can be used to improve the performance of deepPERFECT framework.

Although we used a DL model in this study, other methods including analytical and physical-based models like the finite element method could potentially be used. Previously, these methods have been used to predict the deformation of the body and organs because of surgical procedures and physiological deformation[4–6,14]. The downside of the DL models is that they require high computational resources, a large amount of data, and a long training time, however, their main advantage is very short run-time. Our DL model generates the DFs in less than a second, but a finite element model may take hours to do the analysis.

Table 1 shows the results of the quantified image synthesis evaluation. The 3D Pix2Pix with large patches has the best performance among the other structures. Although the 3D Pix2Pix with small patches and 2.5D Pix2Pix require much less GPU memory, they have lower performance. We believe this is because the differences between the dCT and pCT are clearer in a large field of view. The U-Net structure has the worst performance compared to GAN models, which demonstrates the superiority of GAN models in the synthesis task. Here, we generated DFs, therefore the CT intensity of the sCT remains undistorted and calibrated, as a result, the sCT can directly be used for planning. The average DSC for GTV contour was 0.82. The lowest GTV DSC was observed for cases with the maximum delay. We also observed a trend between DSCs and delay, the longer the delay, the lower the DSC. In addition, the GTV location is subjected to abdominal day-to-day movement and filling which makes the exact prediction of the GTV location nearly impossible. In daily clinical practice, this challenge is addressed by aligning the GTV or fiducial markers inside the GTV under the guidance of onboard cone beam CT and using adaptive radiation therapy using the treatment Day1 CT. [40,41]

Figure 6 shows the bar plots for the V22 Gy and V33 Gy of duodenum, stomach, bowel, and the V33Gy target coverage. The reason V22Gy and V33Gy were chosen is that as seen in the planning protocol these indices are the main clinical constraints for the OARs. Moreover, because the prescribed dose of the SBRT plans was 33Gy, the V33Gy for target volumes is of great interest to the physicians. Our result suggested that the RT plan on the sCT scan has no statistically significant difference from the ground truth. However, when the dose was recalculated on the planning CT, using the planning CT contours, there was a significant difference between the ground truth and recalculated V22Gy and V33Gy. This difference is due to the natural, unpredictable, and patient-specific variation of the abdominal organs. When the isocenter of dose distribution was shifted to the pCT plan isocenter, we achieved full GTV coverage, and although the average PTV coverage was reduced the change in PTV coverage was not statistically significant from the ground truth.

We are aware that our study may have a few shortcomings. First, here we used the small training data. We tried to overcome the issue by using multiple data augmentation methods. As explained in the methods section, we could increase the amount of data from 15 cases to more than 6000 cases. Another shortcoming is that due to GPU limitations, we had to reduce the resolution of the CT images. The loss of resolution in our current model may result in some uncertainty in tumor delineation. The high-quality dCT scan is ideal for tumor delineation due to the high resolution and contrast. As a result, the performance of our current model has degraded due to this loss of quality and resolution. The current resolution may not be enough for clinical treatment planning and delivery as well, however, here our goal was to demonstrate the feasibility of the rapid workflow and as part of future studies, we train the model on high-end GPUs with no degradation in image quality, therefore, since deepPERFECT generates the sCT directly from dCT, the sCT will be a high-resolution scan as well.

Another limitation of our study is that in our current model, we do not include any breathing information. Here, we are using the diagnostic CT to generate the breath-hold planning CT. Considering the patient may have a different breath-hold level, the current model may not be able to capture the patient-specific pattern. To tackle this, we used conditional-GAN model to including patient-specific information by using. A potential remedy is to incorporate a breathing measuring device to measure the patient's breath-hold level by a quick breathing test. Based on this, we will scale the deformation vector fields to match the breating level on the sCT to the patient-specific breath-hold level. However, at the moment this data is not available to us, therefore as part of our future work, to implement this model in our clinical practice, we will measure patients breath-hold level by using spirometer (such shown in the following figure) to create a breathing level aware system.

New-onset of diabetes and weight loss are common features of pancreatic cancer[43]. The severity, extent to which patients are affected, and which type of patients in most affected are not well understood. Studies have reported that patients with high body mass index (BMI), and obesity show the most weight loss[44]. Although we are only using data from 25 patients, the result is consistent with what we have seen in our data. More importantly, another important factor is the time delay between dCT and pCT scans. Patients with low BMI on dCT scan, even with a long delay between two scans, do not show noticeable weight loss on pCT scan. However for patients with almost similar delay, the higher the BMI, the more noticeable change in the patient's body shape is seen. Finally, for one patient with a high BMI, and 2 days delay between two scans, there was no weight loss seen. As a result, although the current model does not account for the weight loss which is reflected in body contour DSC of 0.95, ultimately, with the implementation of deepPERFECT that results in reducing the delay to a few days, the effect of weight loss will be neglectable.

Finally, in the current study, we only applied our DL model to pancreatic cancer cases. However, deepPERFECT can be used for other anatomical sites as well. Our preliminary evaluations showed that our model can successfully be applied to liver and lung cancer patients, but further training and testing are required. We also applied our model to the prostate cancer patient, but due to the considerable differences in the CT field of view, the model showed inferior performance. Nevertheless, the concept of planning CT-free workflow can still be applied to prostate cancer. Therefore, future studies aim at improving the performance of the model by using more data and increasing the resolution of the model using higher-performance GPU resources, and extending the application of deepPERFECT to more anatomical sites.

## Conclusion

We demonstrated the feasibility of planning CT-free rapid pancreatic RT workflow using our deepPERFECT method. We used a fully convolutional 3D/4D GAN DL model to synthesize the planning CT from the initial diagnostic CT. The synthesized CT is compatible with the treatment room setup and mimics the patient shape on the treatment couch. Using this method, we showed that a comparable RT plan to the planning CT plan can be designed on synthesized CT that in turn considerably expedites the workflow and reduces the highly undesirable wait time before RT treatment delivery. This allows the physicians to assess the potential RT outcome well ahead of RT treatment course and verify and adapt the initial plan, if needed, to the treatment day-one CT.

## Acknowledgment

This work was supported by the National Cancer Institute of the National Institutes of Health [award number R37CA229417]. The content is solely the responsibility of the authors and does not necessarily represent the official views of the National Institutes of Health.